\documentclass{ws-ijmpe}

\begin{document}

\markboth{Authors' Names}{Instructions for  
Typing Manuscripts (Paper's Title)}

%%%%%%%%%%%%%%%%%%%%% Publisher's Area please ignore %%%%%%%%%%%%%%%
\catchline{}{}{}{}{}
%%%%%%%%%%%%%%%%%%%%%%%%%%%%%%%%%%%%%%%%%%%%%%%%%%%%%%%%%%%%%%%%%%%%

\title{COALESCENCE OF TWO EXCEPTIONAL POINTS IN THE ANTI-HERMITIAN 3-LEVEL PAIRING MODEL}

\author{\footnotesize J. DUKELSKY}

\address{Instituto de Estructura de la Materia, CSIC, Serrano 123, 28006 Madrid, Spain
\\
dukelsky@iem.cfmac.csic.es}

\author{J. OKO{\L}OWICZ}

\address{Institute of Nuclear Physics, Radzikowskiego 152, PL-31342 Krak\'ow, Poland\\
Jacek.Okolowicz@ifj.edu.pl}

\author{M. P{\L}OSZAJCZAK}

\address{Grand Acc\'{e}l\'{e}rateur National d'Ions Lourds (GANIL), CEA/DSM -- CNRS/IN2P3, \\
BP 5027, F-14076 Caen Cedex 05, France\\
ploszajczak@ganil.fr}

\maketitle

\begin{history}
\received{(received date)}
\revised{(revised date)}
\end{history}

\begin{abstract}
The formation of a higher-order singularity by the coalescence of two exceptional points is studied in the anti-hermitian limit of the complex-extended 3-level Richardson-Gaudin model
\end{abstract}

\section{Introduction}

An essential part of the motion of short-lived nucleonic matter is in
classically forbidden regions and, hence, its properties are effected by both the continuum and many-body correlations \cite{[Oko03],Dob07}. The effect of resonances and the non-resonant scattering states can be considered in the OQS extension of the shell model (SM), the so-called
continuum shell model  (CSM) \cite{[Oko03]}. Two realizations of the CSM have been studied recently: the real-energy CSM \cite{SMEC} and the  complex-energy CSM \cite{Bet02,Mic02}, the so-called Gamow Shell Model (GSM). For hermitian Hamilton operators, both real-energy CSM  and complex-energy CSM (GSM)  lead to the complex-symmetric, non-hermitian eigenvalue problem. As a result, OQSs exhibit unintuitive properties, qualitatively different from  closed quantum systems (CQSs). 

One of them is the phenomenon of sharp crossing of resonances with same quantum numbers (same symmetries). Among degeneracies associated with avoided crossings in quantal spectra, one finds a diabolic point (DP) \cite{Ber84}, and  an exceptional point (EP) \cite{Zir83}, which appears in complex extended hermitian Hamiltonians. In this Letter, we shall discuss some aspects of degeneracies in spectra of a prototypical OQS, the 3-level Richardson-Gaudin (RG) model.

\section{The model}

RG models \cite{Duk04} are based on the $SU(2)$ algebra with elements $K^+_l$, $K^-_l$, and 
$K^0_l$, fulfilling the commutation relations: $[K^+_l,K^-_{l^{\prime}}]=\delta_{ll^{\prime}} K^0_l~$ ,
$[K^0_l,K^{\pm}_{l^{\prime}}]=\pm\delta_{ll^{\prime}} K^{\pm}_l~$, where indices 
$l,l^{\prime}$ refer to a particular copy from a set of $L$, $SU(2)$ algebras.
In the following, we shall use the pair representation of SU(2) algebra. The elementary operators in this representation are the number operators $N_j$ and the pair operators $A^{\dagger}_j$, $A_j$, defined as:
\begin{eqnarray}
N_{j}=\sum_{m}a_{jm}^{\dagger}a_{jm} ~;~~~~~
A_{j}^{\dagger}=\sum_{m}a_{jm}^{\dagger}a_{j\overline{m}}^{\dagger}=(A_{j}^{\dagger})^{\dagger}%
\label{eq2}
\end{eqnarray}
where $j$ is the total angular momentum and $m$ is the $z$-projection. The state 
${j\overline{m}}$ is the time reversal of ${jm}$. The relation between the operators of the 
pair algebra and the generators of the SU(2) algebra is:
\begin{eqnarray}
K_{l}^{0}=\frac{1}{2}N_{l}-\frac{1}{4}\Omega _{l}~;~~~~  K_{l}^{+}& =\left( K_{l}^{-}\right) ^{\dagger
}=\frac{1}{2}A_{l}^{\dagger } \label{K0}
\end{eqnarray}
where $\Omega_l$  is the particle degeneracy of level $l$. With this correspondence, one 
can introduce the non-integrable 3-level pairing Hamiltonian:
\begin{eqnarray}
H(g)=\sum_{i}\varepsilon_{i}N_{i}+g\sum_{ij}A_{i}^{\dagger}A_j+g'\sum_{i}N_i^2~~~~, ~~~~ g'=\gamma g
\label{NIPH}
\end{eqnarray}
where $\gamma<0$ is a $C$-number and coupling constants $g$, $g'$ are complex. For $\gamma=0$, the Hamiltonian (\ref{NIPH}) is integrable.

\subsection{Level degeneracies}

The position of degeneracies in the complex g-plane are indicated by the roots of coupled equations:
\begin{eqnarray}
{\rm det}\left[H\left(  g\right)  -EI\right]  = 0~;~~~~ \frac{\partial}{\partial E} {\rm det}\left[ H\left(
g\right)  -EI\right] = 0
\end{eqnarray}
Eliminating $E$ from these two equations, one is left with a polynomial in $g$ of degree  $M\leq n(n-1)$.
The maximal degree of $D(g)$ ($M=n(n-1)$) appears for non-integrable Hamiltonians.
The eigenvalue degeneracies $E_m(g)=E_{m^{\prime}}(g)$ at $g=g_{\alpha}$ ($\alpha=1,\dots,M$), 
can be found numerically by looking for sharp minima of $D(g)$ in the complex $g$-plane. Degenerate eigenvalues can be single-root (EP) or  double-root solutions of $D(g)$. 
%An accidental degeneracy of two states in the trivial case $g=0$, may also be a double-root solution.

\section{Formation of the pseudo-diabolic point in the anti-hermitian Hamiltonian}

In the following, we shall analyze solutions of the 3-level non-integrable pairing model (\ref{NIPH}) in the case of 2 pairs of fermions in a valence space of degeneracy 
$\Omega_1=2, \Omega_2=6, \Omega_3=2$. Energies of levels are 
$\varepsilon _{1}=0$, $\varepsilon _{2}=1$, and $\varepsilon _{3}=2$. In this model space, there are $n=4$ many-body states, and the total number of roots of $D(g)$ is $M=12$. The  3-level pairing model is reflection symmetric with respect to the real $g$-axis  (${\cal I}m(g)=0$). In the lower half-plane, all eigenvalues are either discrete states at ${\cal R}e(g)=0$ or decaying resonances. Complex conjugate states are situated in the upper half-plane (${\cal I}m(g)>0$) and correspond to capturing resonances. 
%%++ Figure:1
\begin{figure}[t]\centering
\includegraphics[width=60mm,angle=-90]{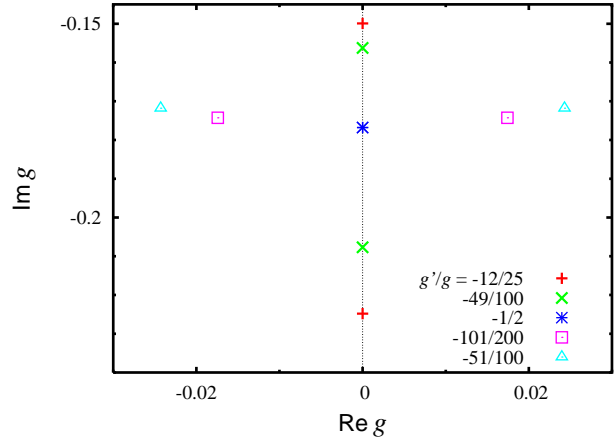}
\caption{The formation and decay of a pseudo-DP at $g'=-g/2$ (the blue star) in the anti-hermitian limit of the non-integrable 3-level pairing Hamiltonian (\ref{NIPH}).}
\label{fig23}
\end{figure}
%
%%++ Figure:2
\begin{figure}[t]\centering
\includegraphics[width=58mm,angle=-90]{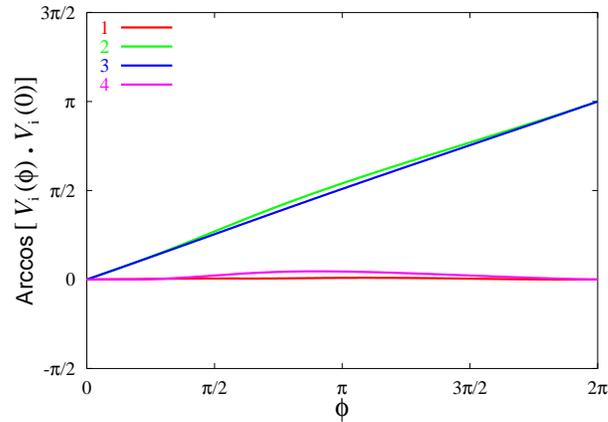}
\caption{Real part of phases for all eigenvectors along the trajectory surrounding the pseudo-DP at $g=-{\rm i}/(4\sqrt{2})$. The phase of each vector is defined  with respect of its reference value at $\phi =0$. For more details, see the discussion in the text.} 
\label{fig24}
\end{figure}
%
%%++ Figure:3
\begin{figure}[t]\centering
\includegraphics[width=53mm,angle=-90]{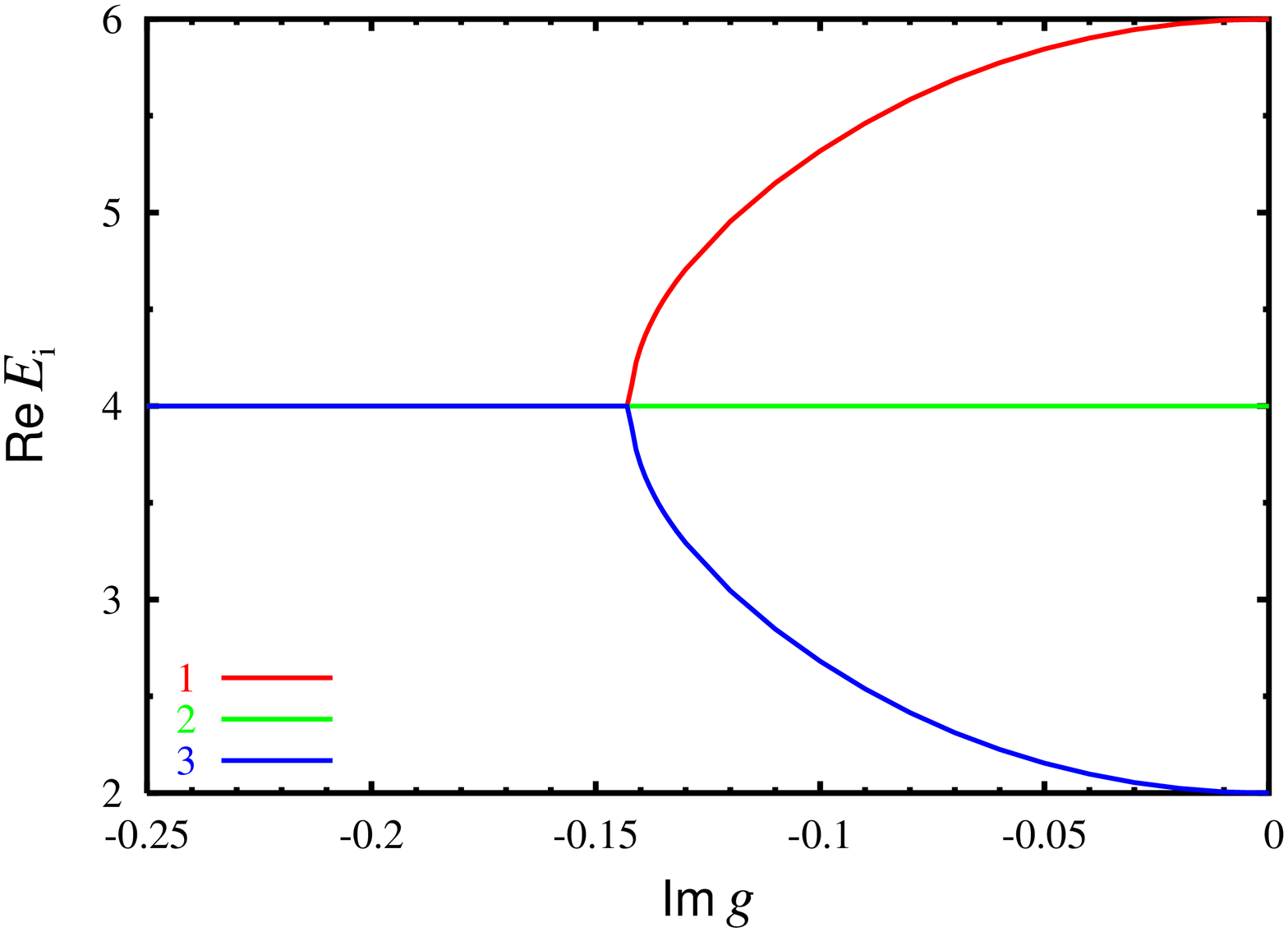}
\includegraphics[width=53mm,angle=-90]{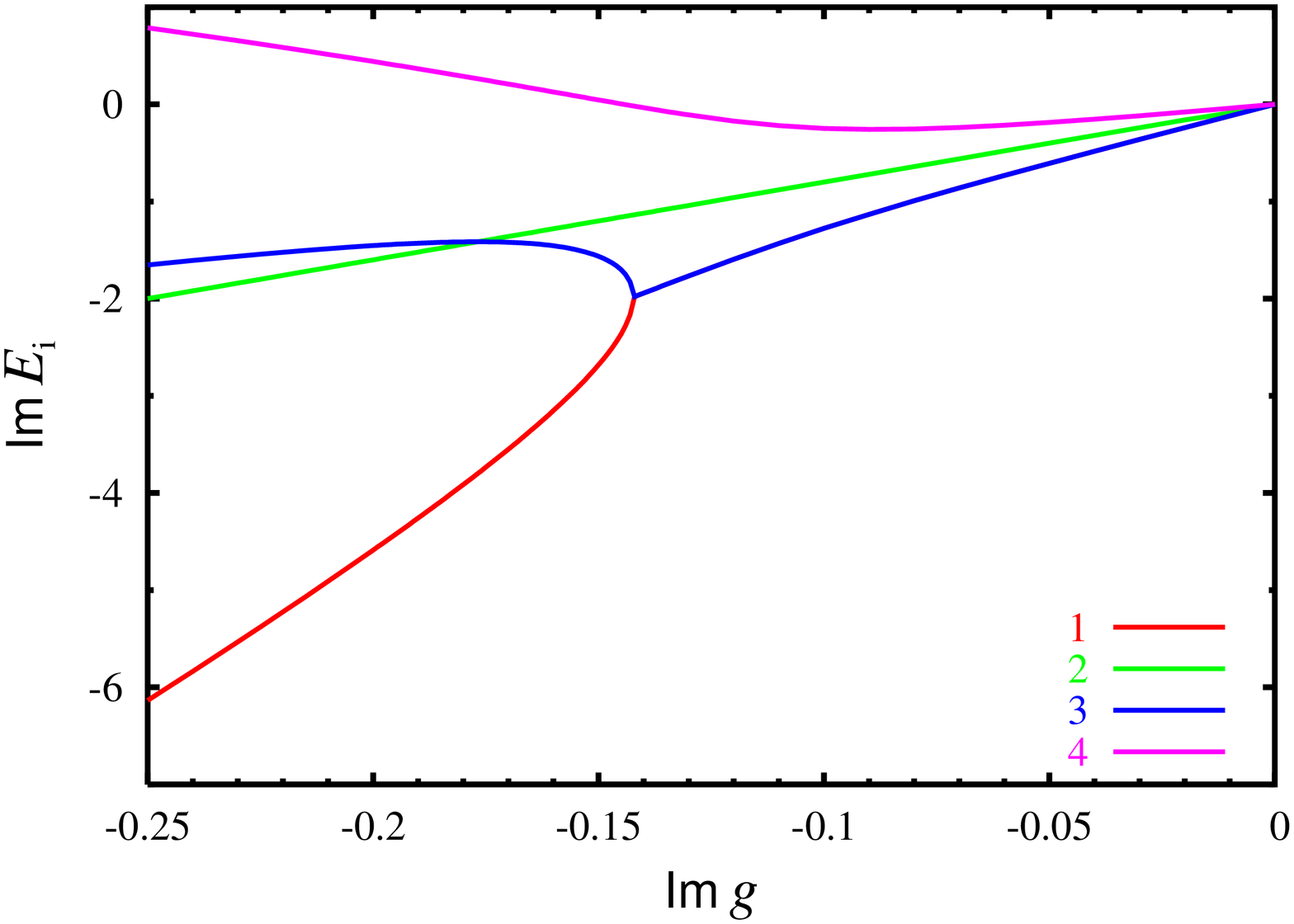}
\caption{Real and imaginary parts of eigenvalues at around the pseudo-DP 
$g=-{\rm i}/(4\sqrt{2})$ are plotted along the cut 
$(0,{\cal I}m(g))$ in $g$-plane. For more details, see the description in the text.}
 \label{fig25}
\end{figure}

Fig. \ref{fig23} shows a pattern associated with the formation/breakup of the pseudo-DP at 
$g=-{\rm i}/(4\sqrt{2})$, {\em i.e.} in the non-hermitian limit of the model (\ref{NIPH}). With an increasing value of $g'/g$, the two EPs in the complex $g$-plane approach each other, coalesce at $g'/g=-1/2$ and move along the axis ${\cal R}e(g)=0$ for $g'/g>-1/2$. At $g=-{\rm i}/(4\sqrt{2})$ for $g'/g=-1/2$, real parts of all eigenenergies are identical. 

Fig. \ref{fig24} shows a real part of the phase along the trajectory in $g$-plane surrounding the pseudo-DP, for all eigenvectors in the chosen model space.  The phase of each vector is defined with respect of its value at $\phi =0$. The trajectory is defined by: 
$g(\phi) = -(1/(4\sqrt{2}){\rm i} + 0.01{\rm e}^{{\rm i}\phi}$. 
The (complex) eigenvalues for vectors '2' and '3' are degenerate at the pseudo-DP and change their phase by $\pi $ at each encircling, as expected for an ordinary DP. However, in contrast to the ordinary DP, both eigenvalues {\em and} eigenvectors merge forming a Jordan block, like for an ordinary EP. Nevertheless, the crossing of eigenvalues is sharp as in the case of ordinary level crossing or the DP. One can see in Fig. \ref{fig24} that phases of  vectors '1' and '4', which have different imaginary parts of eigenvalues, remain approximately constant. 
%Hence,  vectors '1' and '4' remain spectators of the DP formation. 
%It should be stressed that the singularity at  $g=-{\rm i}/(4\sqrt{2})$ is the first to our knowledge case of a DP found in a non-hermitian system.

The complex behavior of eigenvalues in the neighborhood of a pseudo-DP in complex $g$-plane are shown in  Fig. \ref{fig25} along the cut $(0,{\cal I}m(g))$. In the lower part, one can see the appearance  of a pseudo-DP at  $g=-{\rm i}/(4\sqrt{2})\simeq -0.177{\rm i}$ as  a crossing of imaginary parts of eigenvalues '2' and '3'. At 
$g\simeq -{\rm i}/(4\sqrt{2})$, real parts of all eigenvalues are equal (see the upper part of Fig. \ref{fig25}). Notice at $g=-0.142276{\rm i}$, close to the pseudo-DP, a formation of an  EP by eigenvalues '1' and '3'.  

\section{Outlook}
In contrast to a CQS (a hermitian system), the {\em exact} degeneracies of eigenvalues for states of the same quantum numbers (symmetries) are allowed in an OQS. Such degeneracies (EP, DP, pseudo-DP) are closely related to topological anomalies of the Hilbert space \cite{Ber84,Lau94}. In this Letter, we have demonstrated the formation of a pseudo-DP in the anti-hermitian limit of a 3-level pairing Hamiltonian. It remains to be seen in future studies using the CSM and realistic two-body interactions whether those anomalies can have observable consequences in resonance spectroscopy and low-energy nuclear reactions. 

\section*{Acknowledgements}

This work was supported in part by the Spanish MEC under grant No. FIS2006-12783-C03-01 and by the CICYT(Spanish)-IN2P3(French) cooperation.

\end{document}